\documentclass[%
 reprint,longbibliography,preprintnumbers,
nofootinbib,
 amsmath,amssymb,
 aps,
prb,
]{revtex4-2}
\pdfoutput=1
\usepackage[utf8]{inputenc}
\usepackage{flushend}
\usepackage{dcolumn}
\usepackage{bm}
\usepackage{balance}

\usepackage[normalem]{ulem}

\usepackage[colorlinks = true,
            linkcolor = blue,
            urlcolor  = blue,
            citecolor =green,
            anchorcolor = blue]{hyperref}
\usepackage{verbatim}
\usepackage{color,ulem}
\usepackage[english]{babel}

\usepackage[utf8]{inputenc}
\input Starburst.fd
\newcommand*\initfamily{\usefont{U}{Starburst}{xl}{n}}\initfamily

\newcommand{\beq}{\begin{eqnarray}}
\newcommand{\eeq}{\end{eqnarray}}
\usepackage{amsmath}
\usepackage{tikz}
\usetikzlibrary{decorations.pathmorphing}
\usetikzlibrary{shapes.misc}
\tikzset{cross/.style={cross out, draw=black, minimum size=8*(#1-\pgflinewidth), inner sep=0pt, outer sep=0pt},
cross/.default={1pt}}
\usetikzlibrary{patterns,math}
\begin{document}

\title{Understanding the thickness-dependent dielectric permittivity of thin films}

\author{\textbf{Alessio Zaccone}$^{1,2}$}%
 \email{alessio.zaccone@unimi.it}
 
 \vspace{1cm}
 
\affiliation{$^{1}$Department of Physics ``A. Pontremoli'', University of Milan, via Celoria 16,
20133 Milan, Italy.}
\affiliation{$^{2}$Institute for Theoretical Physics, University of G\"ottingen, Friedrich-Hund-Platz 1, 37077 G\"ottingen, Germany}

\begin{abstract}
The dielectric properties of thin films are of paramount important in a variety of technological applications, and of fundamental importance for solid state research. In spite of this, there is currently no theoretical understanding of the dependence of the dielectric permittivity on the thickness of thin films. We develop a confinement model within the Lorentz-field framework for the microscopic Langevin-equation description of dielectric response in terms of the atomic-scale vibrational modes of the solid. Based on this, we derive analytical expressions for the dielectric permittivity as a function of thin film thickness, in excellent agreement with experimental data of Barium-Strontium-Titanate (BST) thin films of different stoichiometry. The theory shows that the decrease of dielectric permittivity with decreasing thickness is directly caused by the restriction in $k$-space of the available eigenmodes for field-induced alignment of ions and charged groups. 
\end{abstract}

\maketitle
\section{Introduction}
The physical properties of thin films are crucial for a variety of technological applications, ranging from optical mirrors to solar cells \cite{solar_science,Friend}, and of fundamental interest in physics. In particular, understanding the dielectric properties of solid thin films is vital in optics, e.g. dielectric (Bragg) mirrors, thin-film interference (anti-reflective coatings), optical and protective coatings, microwave devices, memory devices and 5G wireless communication. Dielectric and electrical properties may strongly depend on thin film thickness, which is a problem of both fundamental and technological interest \cite{Parker,Natori,Starkov_2016}. In particular, size effects give rise to changes in the electrical performance of thin film capacitors and field-effect transistors, including issues such as depolarization fields in the dielectric sandwiched between semiconductors \cite{Silverman}, and polarization screening in metal-dielectric-metal thin-film capacitors \cite{IEEE}. Finally, thin films are known to have enhanced or ultra-high dielectric strength, which is another important thickness-dependent effect \cite{dielectric_strength}.
In spite of these tremendous technological implications, there exists currently no quantitative theoretical description of the dependence of dielectric properties on film thickness. This is also due to the intrinsic limitations of ab initio methods which cannot simulate thicknesses of more than a few nanometers \cite{Yim2015,Stengel2006}. We provide here for the first time a microscopic theory able to quantitatively describe this effect.

\section{Theory}
\subsection{Langevin equation framework for lattice ions}
We describe the dynamics of charged groups (ions), and of partially-charged groups, in a solid material by means of a generalized Langevin equation (GLE) \cite{Coffey,Cui_GLE}. Since we are interested in determining the dielectric response of the material, we have to use a GLE that contains the appropriate Lorentz force term due to the external oscillating electric field, and which contains a stochastic force term that obeys a suitable fluctuation-dissipation theorem. Denoting the mass-rescaled tagged-particle displacement $s=Q\sqrt{M}$ where $M$ is the mass of the ion (or of the partially charged group) and $Q$ the displacement vector, the resulting equation of motion, reads as \cite{Cui_GLE}
\begin{equation}
\ddot{s}=-\mathcal{U}'(s)-\int_{0}^t\nu(t-t')\frac{ds}{dt'}dt'+F(t)+zE(t),
\end{equation}
where $U(s)$ denotes the local force field, $\nu$ is the microscopic friction due to the long-range non-local anharmonic interactions with the thermal bath represented by all other atoms and ions present in the system. Furthermore, $F(t)$ is the stochastic force representing the thermal noise, and the last term on the RHS is the Lorentz force term relating to the system's response to the external AC electric field, where the charge $z$ has been redefined to be the mass-scaled charge. In order to determine the dependence of the polarisation and of the dielectric function on the frequency of the field in 3D space, we have to describe the displacement $\mathbf{s}$ of each charged particle from its own equilibrium position under the applied AC field $\mathbf{E}(t)$. Upon treating the dynamics classically, the equation of motion for a charge $I$ under forces coming from interactions with other atoms in the system and from the applied AC electric field is given by
\begin{equation}
\label{3.3gle}
\ddot{s}_I^\mu=-\sum_{J\nu}H_{IJ}^{\mu\nu}s_J^\nu-\int_{0}^t\nu(t-t')\frac{ds_I^\mu}{dt'}dt'+F_{I}^\mu(t)+z_I E^\mu(t).
\end{equation}
under the assumption of pairwise interactions and the Greek index $\mu$ denotes space components of a vector.

The next step is to take the Fourier transform, $s_I^\mu(t)\rightarrow\tilde{s}_I^\mu(\omega)$, leading to 
\begin{equation}
-\omega^2\tilde{s}_I^\mu(\omega)+i\omega\tilde{\nu}(\omega)\tilde{s}_I^\mu(\omega)+H_{IJ}^{\mu\nu}\tilde{s}_J^\mu(\omega)=\tilde{F}_{I}^\mu+z_I\tilde{E^\mu}
\end{equation}
where the tilde is used to indicate Fourier-transformed quantities. Hence, $\tilde{\nu}(\omega)$ denotes the Fourier transform of $\nu(t)$.
Since the Hessian matrix $H_{IJ}^{\mu\nu}$ is real and symmetric and its eigenvectors provide a basis set in Hilbert space, we can apply normal-mode decomposition by projecting the $3N$-dimensional Fourier-transformed displacement vector $\tilde{s}$ onto the $3N$-dimensional eigenvectors of the Hessian: $\hat{\tilde{s}}_m(\omega)=\tilde{s}(\omega)\cdot e_{m}$. Here the hat is used to denote the coefficient of the projected quantity, $e_{m}$ represent orthonormal eigenvectors of the Hessian matrix, and $m$ runs from $1$ to $3N$. Then we obtain
\begin{equation}
-\omega^2\hat{\tilde{s}}_m+i\omega\tilde{\nu}(\omega)\hat{\tilde{s}}_m+\omega_m^2\hat{\tilde{s}}_m=\hat{\tilde{F}}_m+(z\hat{\tilde{E}})_m.
\end{equation}
The equation is solved by
\begin{equation}
\hat{\tilde{s}}_m(\omega)=-\frac{\hat{\tilde{F}}_m+(z\hat{\tilde{E}})_m}{\omega^2-i\omega\tilde{\nu}(\omega)-\omega_m^2}.
\end{equation}
Upon transforming back to a vector equation for the Fourier-transformed
displacement of charge $I$, we have:
\begin{equation}
\sum_{I}\tilde{\mathbf{s}}_I(\omega)=\sum_{m}-\frac{\tilde{\mathbf{F}}+z\tilde{\mathbf{E}}}{\omega^2-i\omega\tilde{\nu}(\omega)-\omega_m^2},
\label{3.3sI}
\end{equation}
where $\tilde{\mathbf{F}}$ and $\tilde{\mathbf{E}}$ are average values.

\subsection{Polarization and dielectric function}
Each charged particle contributes to the polarisation a moment $\mathbf{p}_I=z_I\mathbf{s}_I$. In order to evaluate the macroscopic polarisation, we add together the averaged contributions from all microscopic degrees of freedom in the system, $\mathbf{P}=\sum_I\mathbf{p}_I$. In order to do this analytically, we use the fact that the ensemble average of the noise for an oscillating system vanishes upon averaging over many cycles, as demonstrated also numerically in Ref. \cite{Damart}. We thus employ the sum over all contributions of the type given by Eq. \eqref{3.3sI}, to obtain the averaged polarisation. We also perform the standard procedure of replacing the discrete sum over the total degrees of freedom of the solid with the continuous integral over the eigenfrequencies $\omega_m, \sum_m...\rightarrow\int g(\omega_p)...d\omega_p$, where $g(\omega_p)$ is the vibrational density of states (VDOS). This gives
the following sum rule for the electric polarisation \cite{CuiPRE2017}:
\begin{equation}
\tilde{\mathbf{P}}(\omega)\varpropto-\left[\int_0^{\omega_D}\frac{g(\omega_p)}{\omega^2-i\omega\tilde{\nu}(\omega)-\omega_p^2}d\omega_p\right]\tilde{\mathbf{E}}(\omega)
\end{equation}
Here, $\omega_D$ is the cutoff
Debye frequency arising from the normalisation of the VDOS (i.e. the highest eigenfrequency in the vibrational spectrum). Furthermore, we have defined a $3N$-dimensional vector $\vec{z}$ such that $\hat{z}_m=\vec{z}\cdot e_{m}$ is a scalar factor, arising from Eq. \eqref{3.3sI} \cite{cui_2020}, which is later going to be absorbed into the prefactor $A$ and therefore is no longer shown in the above relation.

Note that we have taken an ensemble average over the system. The complex dielectric permittivity $\epsilon$ is defined as $\epsilon=1+4\pi\chi_e$, where $\chi_e$ is the dielectric susceptibility, which relates polarisation and electric field via
$\mathbf{P} = \chi_e\mathbf{E}$ \citep{Born-Wolf}.

Within this model \cite{CuiPRE2017}, the dielectric function is finally expressed as a frequency integral as \cite{CuiPRE2017}
\begin{equation}\label{3.3diele}
\epsilon(\omega)=1-A\int_0^{\omega_D} \frac{g(\omega_p)}{\omega^2-i\omega\tilde{\nu}(\omega)-\omega_p^2} d\omega_p
\end{equation}
where $A$ is an arbitrary positive constant, whose numerical value has to be matched with experiments. Clearly, if $g(\omega_{p})$ were given by a Dirac delta, one would recover the standard simple-exponential Debye relaxation \cite{CuiPRE2017,Froehlich}. This approach can be extended to deal with atoms and molecules that have stronger inner polarisability by replacing the external field field $\mathbf{E}$ with the local electric field $\mathbf{E}_{loc}$, which is known as the Lorentz cavity model or Lorentz field \cite{Froehlich,Choy}.
In condensed matter, the net electric field that acts on a molecule locally is equal to the external field only for vanishing polarisability of the molecule. This is a well-known effect, whereby the field in the medium is influenced (typically, diminished) by the local alignment of the nearby polarised molecules. The simple Lorentz cavity model works well in materials where the building blocks are not pathologically shaped or too anisotropic. In order to keep the treatment analytical, we focus on the case of constant friction, $\nu=const$.
The derivation of the local field or Lorentz field can be found in many textbooks \cite{Froehlich,Choy} and gives
\begin{equation}
\mathbf{E}_{\textit{loc}}=\mathbf{E}+\frac{4\pi}{3}\mathbf{P}.
\end{equation}
Therefore, $\mathbf{E}$ is replaced with the Lorentz field $\mathbf{E}_{\textit{loc}}$, and the equation of motion becomes:
\begin{equation}
\tilde{\mathbf{s}}_{I}^\prime(\omega)=\frac{z_I}{\omega^2-i\omega\nu-
\omega_{p}^2}\left(\tilde{\mathbf{E}}(\omega)+\frac{4\pi}{3}\mathbf{\tilde{P}}(\omega)\right).
\end{equation}

Combining the above relations and summing over all contributions from all the ions and charged groups, we obtain
\begin{align}
&\mathbf{P}=\left(\sum_{I} q_I \mathbf{s}_{I} + \alpha\mathbf{E}_{\textit{loc}}\right);\quad
\epsilon(\omega)=1+4\pi\frac{\chi(\omega)}{1-\frac{4\pi}{3}\chi(\omega)},\notag\\
&\chi(\omega)=\int_0^{\omega_D}\frac{A\,g(\omega_p)}{\omega^2-\omega_p^2+i\omega\nu}d\omega_p+\alpha
\label{lorentz}
\end{align}
where $\alpha$ is the microscopic electronic polarisability and we used the definition of electric displacement vector, $\mathbf{D}=\epsilon\mathbf{E}=\mathbf{E}+4\pi\mathbf{P}$ \cite{Jackson}. Furthermore, in the expression of $\chi$ we have incorporated the factor $z^2$ into the rescaling coefficient $A$.
This microscopic theory of the dielectric response has been previously applied to describe experimental data of supercooled glycerol in Ref. \cite{CuiPRE2017} and was able to explain the non-Debye asymmetric excess wing of the dielectric loss $\epsilon''(\omega)$ as due to the excess of low-energy vibrational modes that characterizes the vibrational spectra of supercooled liquids and glasses.

\subsection{Thin film confinement and thickness-dependent permittivity}
\begin{figure}[h]
\centering
\includegraphics[width=0.9\linewidth]{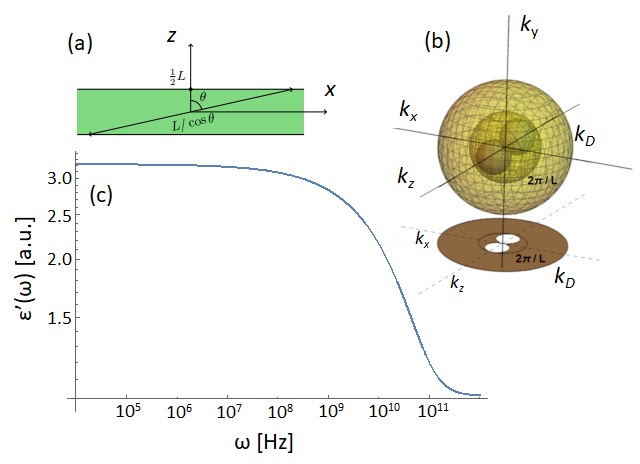}
\caption{Panel (a) shows the thin film geometry in real space (confined along the $z$-axis but unconfined along the $x$ and $y$ axis), with the maximum wavelength that corresponds to a certain polar angle $\theta$. Panel (b) shows the corresponding geometry of $k$-space, where the outer Debye sphere (of radius $k_{D}$) contains two symmetric spheres of forbidden states, i.e. states in $k$-space that remain unoccupied due to confinement along the $z$-axis See Ref. \cite{Phillips} for a detailed mathematical derivation of this result. Panel (c) shows an illustrative calculation of the dielectric permittivity using the confined model, with typical values of speed of sound (in the order of kHz and of microscopic friction $\nu$ and Debye frequency both in the order of $10^{13}$ Hz. Panels (a) and (b) 
 have been adapted from Ref. \cite{Phillips}, with permission from the American Physical Society.}
\label{fig1}
\end{figure}

The above form of the dielectric function is derived for a bulk material. In the case of a thin film, the confinement along the vertical direction imposes a constraint on the wavelength of the vibrational excitations that are allowed to propagate \cite{PNAS2020}. As shown mathematically in Ref. \cite{Phillips}, and verified against MD simulations and experiments for real thin solid films in \cite{Yu_2022}, the thin-film confinement imposes a low-frequency cut-off on the wavevector of vibrational modes equal to $k_{min}=\frac{2\pi \cos \theta}{L}$. Here $L$ is the film thickness and $\theta$ is the polar angle with the vertical $z$-axis, see Fig. \ref{fig1}. 
The condition $k_{min}=\frac{2\pi \cos \theta}{L}$ identifies two spheres of forbidden states inside the Debye sphere.

As demonstrated in Ref. \cite{Yu_2022}, this lower cut-off wavevector corresponds to a minimum frequency of the allowed vibrational modes given by $\omega_{min}=c \,k_{min}$, where $c$ is a characteristic speed of sound (independent of $L$). For example, $c$ can be taken as the average speed of sound $c$ used in Debye's theory, defined as $\frac{3}{c^3}=\frac{1}{c_L^3}+\frac{2}{c_T^3}$, where $c_L$ and $c_T$ are the longitudinal and the transverse speed of sound, respectively
Since the minimum wavevector depends on the polar angle $\theta$, we can take a spherical average over the solid angle that gives the average minimum wavevector $\bar{k}_{min}=\frac{\pi}{L}$, and a minimum frequency $\bar{\omega}_{min}=\frac{c\pi}{L}$. Upon implementing this confinement-induced cut-off in the susceptibility integral we get: 

\begin{align}
&
\epsilon(\omega)=1+4\pi\frac{\chi(\omega)}{1-\frac{4\pi}{3}\chi(\omega)},\notag\\
&\chi(\omega)=\alpha+\int_{\frac{c\pi}{L}}^{\omega_D}\frac{A\,g(\omega_p)}{\omega^2-\omega_p^2+i\omega\nu}d\omega_p. \label{confined}
\end{align}

As a sanity check, we plot in Fig. \ref{fig1}(c) a typical behaviour of the dielectric permittivity $\epsilon'(\omega)$ computed using Eq. \eqref{confined} for realistic values of frequencies encountered in solid materials. The resulting curve still presents all the typical features of dielectric permittivity as a function of frequency, with a low-frequency plateau followed by a drop (dielectric relaxation) in the frequency range $10^{8}-10^{9}$ Hz typically measured in dielectric spectroscopy \cite{Kremer}.

In order to understand the effect of film thickness $L$ on the permittivity $\epsilon'$, the integral in Eq. \eqref{confined} has to evaluated. Using a Debye law for the vibrational density of states, $g(\omega_p) \sim \omega_{p}^{2}$, as appropriate for a solid, the integral cannot be evaluated analytically. Nevertheless, we can approximate the integral for low-to-intermediate frequencies $\omega$ since this is the regime of interest for measurements of dielectric permittivity of thin solid films. Using the approximation $\omega_p \gg \omega$, for the real part of the integral we obtain:
\begin{equation}
    \int\frac{\omega_p^4}{\omega_p^4+\omega^2\nu^2}d\omega_p \approx \omega_p -\sqrt{\frac{\nu\,\omega}{2}}\arctan\left(\frac{\sqrt{2} \,\omega_p}{\sqrt{\nu\,\omega}}\right).
\end{equation}
With the integration limits set in Eq. \eqref{confined}, and neglecting nonlinear contributions to the susceptibility, this leads to the following expression for the thickness-dependent dielectric permittivity:
\begin{widetext}
\begin{align}   \epsilon'(\omega)&=\epsilon_{\infty}+ 
4\pi A\frac{  (\omega_D -\frac{c\pi}{L} )+\sqrt{\frac{\nu\,\omega}{2}}\arctan\left(\frac{\sqrt{2}}{\sqrt{\nu\,\omega}}\frac{c\pi}{L}\right)- B}{1-\frac{4\pi}{3}A(\omega_D- \frac{c\pi}{L}+\sqrt{\frac{\nu\,\omega}{2}}\arctan\left(\frac{\sqrt{2}}{\sqrt{\nu\,\omega}}\frac{c\pi}{L}\right)- B)},\nonumber \\
B&=\sqrt{\frac{\nu\,\omega}{2}}\arctan\left(\frac{\sqrt{2} \,\omega_D}{\sqrt{\nu\,\omega}}\right). \label{thickness}
\end{align}
\end{widetext}
We note that, in the limit $L \rightarrow \infty$, the above expression correctly tends to a constant value independent of $L$, which represents the bulk value at a given frequency $\omega$.
The above expression Eq. \eqref{thickness} can be Taylor expanded in $L$ to study the leading terms that control the thickness-dependent dielectric permittivity of thin films. To second order in $L$ we thus obtain:
\begin{equation}
\epsilon'(\omega)=\epsilon_{\infty} + K_1\, L - K_2\, L^2 + \emph{O}(L^3).\label{expansion}
\end{equation}
where 
\begin{equation}
K_1 \approx  \frac{9-3\,B}{A'}, ~~~~K_2 \approx \frac{  4\pi\,A^2 \sqrt{\nu \omega}\,  \omega_D}{A'^2},\label{constants}
\end{equation}
where $A'\equiv 4 \pi^2 A\, c$ has units of length, which is reassuring because then in Eq. \eqref{expansion} all terms are dimensionless, as they should.

\section{Comparison with experimental data}
For thin film oxides at low to intermediate frequencies $\omega = 1-10$ kHz, one has $\epsilon' \sim 10-100$ from experimental measurements. Then it is clear that the rescaling constant $A$, which, in the above, multiplies the Debye frequency $\omega_D \sim 10^{13}$ Hz, must be in the order of $A \sim 10^{-12}$, hence, in general, a small number. From the first one of Eqs. \eqref{constants}, it follows that $K_1 >0$ always, provided that: $\frac{9}{A}>\frac{3B}{A}=3\sqrt{\nu \omega}$. This is always true because, for realistic values of experimental systems, one has $\sqrt{\nu \omega} \sim 10^8 - 10^9$ Hz, which is orders of magnitude smaller than $\frac{3}{A} \sim 10^{12}$ Hz. 
We thus conclude that $K_1$ is always positive, and therefore the leading term in the expansion is such that the dielectric permittivity increases with increasing $L$. 
This means that, overall, the confinement acts as to lower the dielectric permittivity. The above microscopic theory explains that, physically, this is due to the cutting-off of low-frequency vibration eigenmodes at the atomic level due to the confinement, which, in turn, leads to more limited possibilities for the ions and charged groups to rearrange spatially (''align'') in response to the local electric field. 

The second term in the expansion is, instead, negative and acts as to level off the initial increase as a function of thickness $L$. 
Furthermore, with realistic values of the physical parameters as declared above, we have that $K_2 \sim 10^{7}$, since $\sqrt{\nu \omega} \sim 10^8$ Hz and $c \sim 10^4$ m/s, while $K_1 \sim 10^{10}$, and thus $K_2/K_1 \sim 0.001$.

This observation then leaves just one non-trivial fitting parameter in the comparison between Eq. \eqref{expansion} and the experimental data, which is reported in Fig. \ref{fig2} below. The non-trivial fitting parameter is $\epsilon_\infty$, which represents the infinite-frequency limit of the dielectric permittivity and is thus controlled by the atomic-scale physics.

\begin{figure}[h]
\centering
\includegraphics[width=0.98\linewidth]{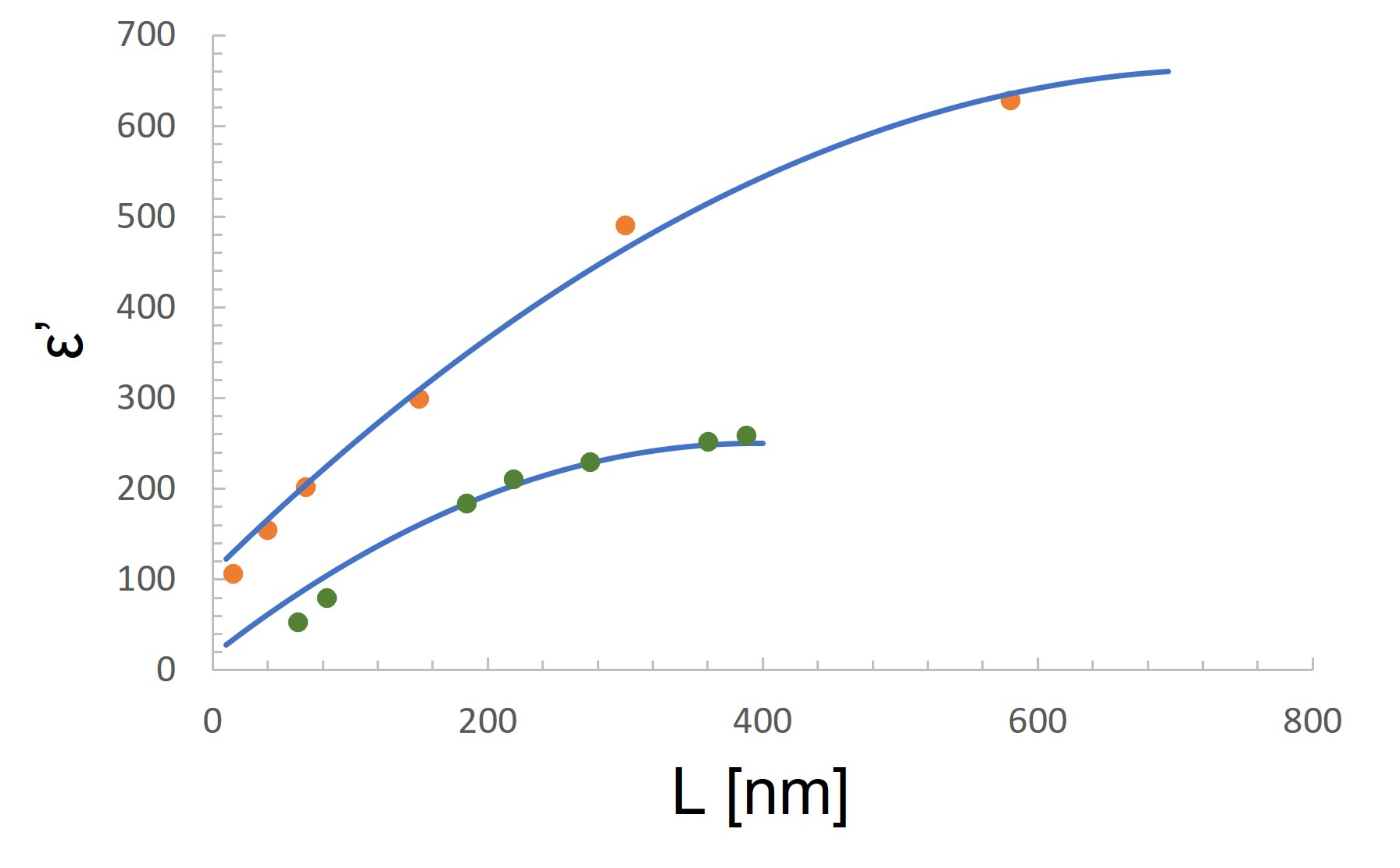}
\caption{Comparison between the theoretical prediction given by Eq. \eqref{expansion}, continuous line, and experimental data (circles). The latter are in arbitrary physical units, as customary for dielectric spectroscopy measurements. The upper curve refers to experimental data of dielectric permittivity of (Ba$_{0.7}$, Sr$_{0.3}$)TiO$_3$ (BST) thin films measured at $\omega = 4$ kHz from Ref. \cite{Parker}, fitted by Eq. \eqref{expansion} with $K_1=1.5$ and $K_2=0.001$, and $\epsilon_\infty=100$. The lower curve refers to experimental data of dielectric permittivity of (Ba$_{0.5}$, Sr$_{0.5}$)TiO$_3$ thin films averaged between $\omega =400$ Hz and $\omega =10$ kHz from Ref. \cite{koreans}, fitted by Eq. \eqref{expansion} with $K_1=1.2$ and $K_2=0.0015$, and $\epsilon_\infty=10$. All experimental measures were made at room temperature.}
\label{fig2}
\end{figure}

Furthermore, the value of $\epsilon_\infty$ is also constrained to be reasonable and much smaller than the bulk value at kHz frequencies, which is indeed the case in the fitting shown in Fig. \ref{fig2}. This further consideration reflects the fact that the above fitting is physically meaningful and reliable.

\section{Conclusion}
In summary, we have developed a microscopic theory of dielectric response of thin solid films starting from a Langevin equation for the motion of charged and partially-charged atoms in the solid layer. Using a recent wave-confinement model, we have adapted the theory to the case of thin films, by implementing a cut-off in momentum space reflecting the fact that a significant population of large-wavelength vibrational modes become forbidden due to the thin-film confinement. In turn, this reduces the possibilities for atomic-scale rearrangements/alignements under the applied field, leading to a lower permittivity for thinner films. The theory leads to an analytical expression for the dielectric permittivity as a function of applied field frequency and film thickness, in excellent agreement with experimental data with just one non-trivial fitting parameter ($\epsilon_{\infty}$), which, however, is constrained to be in a reasonable range by the material physics.
In future work, this theory can be extended to nano-confined liquid films, including nano-confined water \cite{Podgornik,yang2023suppressed}. To this aim, it may be useful to provide a formulation of the above theory also for off-diagonal tensor components. It can also be extended to ultra-thin films (with thickness on the order of few nanometers or lower), where the vibrational density of states features a low-energy $\omega^3$ behaviour, instead of Debye's $\omega^2$ law \cite{Yu_2022}. 
In the future, this theory can open up new ways of tuning and optimizing the electrical performance of thin film devices, ranging from photovoltaics to 5G technology, and of understanding and modelling the ultra-high dielectric strength of thin films \cite{dielectric_strength}.

\subsection*{Acknowledgments} 
The author gratefully acknowledges funding from the European Union through Horizon Europe ERC Grant number: 101043968 ``Multimech'', from US Army Research Office through contract nr.   W911NF-22-2-0256, and from the Nieders{\"a}chsische Akademie der Wissenschaften zu G{\"o}ttingen in the frame of the Gauss Professorship program. Discussion and input from S. Achilli and G. Onida are gratefully acknowledged.

\bibliography{refs}

\end{document}